\documentclass[11pt,a4paper]{article}

\usepackage[utf8]{inputenc}
\usepackage[T1]{fontenc}
\usepackage{lmodern}
\usepackage{geometry}
\usepackage{graphicx}
\usepackage{booktabs}
\usepackage{array}
\usepackage{amsmath,amssymb}
\usepackage{siunitx}
\usepackage{caption}
\usepackage{subcaption}
\usepackage{float}
\usepackage{hyperref}
\usepackage{xcolor}
\usepackage{setspace}

\graphicspath{{figures/}}
\hypersetup{colorlinks=true, linkcolor=blue, citecolor=blue, urlcolor=blue}
\title{A technical report on the surface-energy and morphology-based screening for electrode/electrolyte interface compatibility in SOFC/ReSOC materials}
\author{G. Violano and L. Afferrante\\
\small Department of Mechanics, Mathematics and Management, Politecnico di Bari,\\
\small Via E. Orabona 4, 70126 Bari, Italy}
\date{}

\begin{document}
\maketitle

\begin{abstract}
The performance and durability of solid oxide and reversible solid oxide cells are strongly affected by the electrode/electrolyte interface, where charge transfer, ionic transport, adhesion, morphology and thermomechanical stability interact. However, early-stage compatibility screening is usually based on electrochemical or compositional criteria, while surface-related descriptors are rarely included in a unified framework.

This report proposes a surface-based methodology to assess the expected compatibility of candidate electrode/electrolyte pairings. Contact-angle measurements with water and glycerol are used to determine total, dispersive and polar surface free energy components through the Owens--Wendt--Rabel--Kaelble method. Confocal topography is used to extract ISO 25178 roughness parameters, including average roughness, peak-to-valley height, valley depth, skewness, kurtosis and surface slope.

A compatibility matrix is constructed by combining energetic affinity and morphological suitability, with emphasis on the electrolyte surface, since the electrode is deposited directly onto the electrolyte substrate. The results indicate that the most promising interfaces are not necessarily those with the highest surface free energy, but those combining high adhesion work, low interfacial energy and a substrate morphology suitable for continuous electrode deposition.

The proposed approach provides a rational pre-electrochemical screening tool to prioritize electrode/electrolyte combinations for subsequent validation by electrochemical impedance spectroscopy, area specific resistance, electrical contact resistance, microstructural analysis and durability testing. Although it does not replace electrochemical characterization, it offers a physically grounded way to connect surface chemistry, topography and interface formation in solid oxide cell materials.
\end{abstract}

\section{Introduction}

Solid oxide fuel cells (SOFCs) and reversible solid oxide cells (ReSOCs) are high-temperature electrochemical devices for efficient energy conversion, fuel flexibility and energy storage. In SOFC mode, chemical energy from hydrogen, syngas or light hydrocarbons is converted into electricity and heat; in SOEC mode, the same ceramic architecture can operate reversibly to produce hydrogen or syngas from steam and/or carbon dioxide \cite{minh1993,steele2001,singhal2000,stambouli2002,kakac2007}. Their high operating temperature improves reaction kinetics and system efficiency, but also imposes strict requirements on materials compatibility, interfacial stability and thermomechanical durability \cite{singhal2000,stambouli2002,kakac2007,hajimolana2011,golkhatmi2022,chen2011}.

A solid oxide cell consists of a dense ion-conducting electrolyte between two porous electrodes. Yttria-stabilized zirconia remains the benchmark electrolyte, while doped ceria, lanthanum gallate-based perovskites and proton-conducting oxides are widely investigated for lower-temperature operation \cite{minh1993,steele2001,singhal2000,golkhatmi2022,laguna2012,ebbesen2014}. On the electrode side, Ni-based cermets are commonly used at the fuel electrode, while perovskite and mixed ionic-electronic conducting oxides such as LSM, LSCF, BSCF and related composites are used at the oxygen electrode \cite{atkinson2004,simwonis2000,hauch2008,jiang2008,shao2004,adler2004,yokokawa2008}. In all cases, the cell response is governed not only by the intrinsic properties of each material, but also by the quality of the electrode/electrolyte interface.

The electrode/electrolyte interface is a functional region where ionic transport, electronic transport, gas access, catalytic activity, adhesion and mechanical constraint must coexist. In classical composite electrodes, the electrochemical reaction is concentrated near the triple phase boundary, while in mixed conducting electrodes the reaction zone may extend into the electrode volume. In both cases, performance depends on phase connectivity, porosity, surface exchange, ionic/electronic pathways and interfacial continuity \cite{costamagna1998,virkar2000,wilson2006,jorgensen2010,fu2015,brandon2006,kishimoto2011}. Therefore, the geometrical electrode area does not necessarily correspond to the effective electrochemically active area.

This distinction is particularly relevant when electrodes are directly deposited onto dense electrolyte substrates. In such systems, the electrolyte surface controls the first stage of interface formation. Wettability affects spreading of the electrode ink or slurry, surface free energy influences adhesion tendency, and surface morphology determines whether the deposited layer can conform to asperities, fill valleys and avoid local discontinuities. Defects formed during deposition, drying, binder burn-out or firing may evolve into micro-voids, cracks, local thickness variations or weakly bonded regions \cite{chen2011pf,wang2015crack,beckel2007}.

Interface stability is also affected by thermomechanical mismatch. Differences in thermal expansion coefficient between electrode and electrolyte generate residual stresses after firing and cyclic stresses during operation. These stresses can promote cracking, delamination, loss of contact and progressive degradation \cite{selcuk2001,wang2015crack,boukamp1995,mogensen2019,kim2020,faes2012}. From this perspective, adhesion-related descriptors such as the thermodynamic work of adhesion and the estimated interfacial energy can provide useful preliminary indicators of interfacial affinity. They are not equivalent to fracture toughness or critical energy release rate, but they can support early-stage screening of potentially compatible or critical interfaces \cite{griffith1921,irwin1957,wang2015fracture}.

Surface free energy measurements offer a practical way to quantify surface affinity. Using contact-angle measurements and the Owens--Wendt--Rabel--Kaelble method, the total surface free energy can be separated into dispersive and polar components \cite{young1805,fowkes1964,owens1969,kaelble1970,rabel1971,degennes1985}. This distinction is useful for oxide ceramics, where polar interactions may be linked to surface oxygen species, hydroxyl groups, defects, vacancies, cationic sites and adsorbates. However, high surface free energy alone is not sufficient to define a good interface, since energetic affinity must be considered together with surface morphology and deposition compatibility.

Surface morphology further modifies apparent wetting and contact formation, as described by classical Wenzel and Cassie--Baxter concepts \cite{wenzel1936,cassie1944,quere2008}. In ceramic substrates, roughness is not only an average amplitude but also includes valleys, peaks, slopes, skewness and isolated extreme features. ISO 25178 parameters can therefore help describe whether an electrolyte surface is suitable for continuous electrode deposition.

The relationship between surface properties and electrochemical performance must be interpreted carefully. There is no general law directly linking surface free energy to charge-transfer resistance or area specific resistance (ASR) in SOFC/ReSOC systems. Electrochemical response depends on temperature, atmosphere, reaction kinetics, microstructure, transport pathways, catalytic activity and degradation processes \cite{kakac2007,hajimolana2011,costamagna1998,virkar2000}. Nevertheless, surface energy and morphology may influence the fraction of nominal area that becomes effectively active. A poorly wetted or discontinuous interface can reduce the active-area fraction, increase local current density and contribute to higher effective interfacial resistance. A similar link between surface energy, wettability, intimate contact and interfacial resistance has been reported in other solid-state electrochemical systems, particularly in solid-state battery interfaces \cite{dubey2022,banerjee2020,pervez2020}.

The present report therefore uses surface free energy and topography as potential pre-electrochemical compatibility descriptors, not as direct predictors of cell performance. Contact-angle measurements with water and glycerol are used to determine total, polar and dispersive surface free energy components through the OWRK method. Confocal topography is used to extract ISO 25178 roughness parameters. These data are then combined to calculate $W_{12}$ and $\gamma_{12}$ for selected electrode/electrolyte pairings and to build a surface-based compatibility matrix.

The objective is to identify material combinations that are more likely to form continuous, adherent and morphologically suitable electrode/electrolyte interfaces. The resulting ranking is intended to guide subsequent validation by electrochemical impedance spectroscopy, area-specific resistance or electrical contact resistance measurements, microstructural analysis and durability testing. It does not replace electrochemical characterization, but provides a rational preliminary basis for selecting the most promising interfaces for further investigation.

\section{Materials and Methods}

\subsection{Materials and surface-characterization strategy}

The analysed materials include two electrode-side oxides, PBC-HEO and LSFC, and four electrolyte-side oxides, 4PrGDC, 4PrSDC, CZHeLS and CZYbEL. LSFC is a La--Sr--Fe--Co--O perovskite-type electrode material, while 4PrGDC and 4PrSDC are Pr-containing doped-ceria electrolytes based on Ce--Gd--Pr--O and Ce--Sm--Pr--O systems, respectively. PBC-HEO, CZHeLS and CZYbEL are multicomponent oxide formulations.

\begin{figure}[H]
    \centering
    \includegraphics[width=\textwidth]{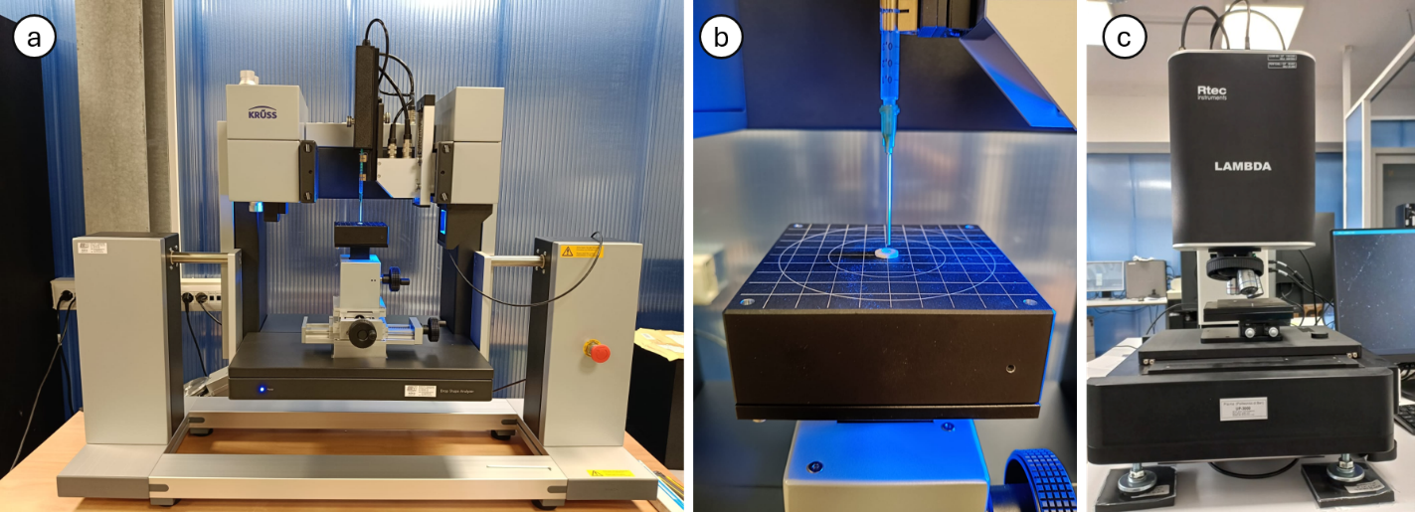}
    \caption{Experimental workflow adopted for surface-based electrode/electrolyte compatibility screening. (a) Optical drop-shape analysis system used for contact-angle and SFE measurements; (b) detail of controlled droplet deposition on the sample surface; (c) confocal microscopy system used for three-dimensional surface-topography acquisition.}
    \label{fig:workflow}
\end{figure}

The experimental setup is shown in Fig.~\ref{fig:workflow}. Contact-angle measurements are used to determine the surface free energy and its dispersive/polar components by the Owens--Wendt--Rabel--Kaelble method, while confocal topography provides ISO 25178 roughness parameters. The two datasets are combined into a compatibility matrix accounting for both energetic affinity and morphological suitability. This matrix is not intended to directly predict impedance, charge-transfer resistance or ASR/ECR values. Instead, it ranks electrode/electrolyte pairings according to their expected ability to form continuous and adherent interfaces.

\subsection{Contact-angle measurements}

Static contact-angle measurements are performed using an optical drop-shape analysis system (Fig.~\ref{fig:workflow}a). For each material, \SI{8}{\micro\liter} droplets of probe liquid are gently deposited on the sample surface (Fig.~\ref{fig:workflow}b), and the droplet profile is recorded by the optical system. The contact angle is then obtained from the droplet shape and used to describe the wetting response of the solid surface.

Water and glycerol are used as probe liquids because their total, polar and dispersive surface-tension components are known and sufficiently different for OWRK surface-energy decomposition. Three measurements are performed on different regions of each sample, and the reported value is taken as the mean contact angle.

Since the investigated surfaces are not ideally smooth, the measured values are treated as apparent contact angles. For this reason, the surface free energy results are interpreted together with the roughness descriptors obtained by confocal microscopy.

\subsection{Surface free energy evaluation by the OWRK method}

The total surface free energy of each solid is expressed as the sum of dispersive and polar contributions:
\begin{equation}
\gamma_s=\gamma_s^d+\gamma_s^p,
\end{equation}
where $\gamma_s^d$ mainly describes dispersive interactions, while $\gamma_s^p$ accounts for polar and chemically specific contributions.

According to the OWRK method, the interaction between a liquid $l$ and a solid $s$ is described by:
\begin{equation}
\gamma_l(1+\cos\theta)=2\left(\sqrt{\gamma_s^d\gamma_l^d}+\sqrt{\gamma_s^p\gamma_l^p}\right),
\end{equation}
where $\theta$ is the measured contact angle, $\gamma_l$ is the liquid surface tension, and $\gamma_l^d$ and $\gamma_l^p$ are its dispersive and polar components.

By applying this relation to water and glycerol, the solid dispersive and polar components, $\gamma_s^d$ and $\gamma_s^p$, are obtained for each surface. These values are then used as input parameters for the electrode/electrolyte compatibility analysis.

For each material, the wetting envelope is derived from the polar and dispersive components of its surface free energy. It defines, in the $\gamma_l^p$--$\gamma_l^d$ plane, the range of liquid surface-tension components expected to satisfy a selected wetting condition. Here, the wetting envelope is used as an auxiliary descriptor of compatibility with possible liquid deposition media, such as electrode inks or slurries. A surface whose wetting envelope matches the liquid phase used for deposition is expected to promote better spreading and a more continuous initial coating.

Thus, the wetting envelope is not treated as a direct performance metric, but as a qualitative tool to support the assessment of deposition suitability and surface compatibility.

\subsection{Confocal topography and roughness parameters}

Surface topography is measured by confocal optical microscopy, which reconstructs a three-dimensional height map of the sample surface. ISO 25178 roughness parameters are then extracted from the reconstructed maps. The selected parameters are reported in Table~\ref{tab:iso_params}.

\begin{table}[H]
\centering
\caption{ISO 25178 topographic parameters considered in the surface-morphology analysis.}
\label{tab:iso_params}
\begin{tabular}{lll}
\toprule
Parameter & Definition & Relevance for interface formation \\
\midrule
$S_a$ & Arithmetical mean height & Average surface roughness \\
$S_q$ & Root-mean-square height & Height dispersion \\
$S_z$ & Maximum height & Peak-to-valley amplitude \\
$S_v$ & Maximum pit depth & Deep valleys \\
$S_{sk}$ & Skewness & Prevalence of peaks or valleys \\
$S_{ku}$ & Kurtosis & Sharp or extreme height features \\
$S_{dq}$ & Root-mean-square gradient & Local surface slope \\
\bottomrule
\end{tabular}
\end{table}

For electrode/electrolyte compatibility, the electrolyte morphology is given particular importance because the electrode is deposited directly onto the electrolyte surface. The electrolyte topography therefore controls initial spreading, coverage and possible filling of valleys by the electrode precursor.

The roughness of the free electrode surface is also considered, but mainly as an indicator of coating quality and homogeneity, rather than as a direct representation of the buried electrode/electrolyte interface.

\subsection{Pairwise electrode/electrolyte surface descriptors}

For each electrode/electrolyte pair, the thermodynamic work of adhesion is estimated using the OWRK/Fowkes expression:
\begin{equation}
W_{12}=2\left(\sqrt{\gamma_1^d\gamma_2^d}+\sqrt{\gamma_1^p\gamma_2^p}\right),
\end{equation}
where subscripts 1 and 2 refer to the two materials forming the interface. Higher $W_{12}$ values indicate a stronger thermodynamic tendency toward adhesion.

The estimated interfacial energy is calculated as:
\begin{equation}
\gamma_{12}=\gamma_1+\gamma_2-W_{12},
\end{equation}
where $\gamma_1$ and $\gamma_2$ are the total surface free energies of the two materials. Lower $\gamma_{12}$ values indicate a more energetically favourable interface.

These quantities are used as energetic descriptors of interface compatibility. They are not treated as direct measures of fracture toughness or electrochemical resistance, but are combined with morphological descriptors to build a relative compatibility ranking.

\subsection{Integrated compatibility index}

An integrated compatibility index is defined to rank the electrode/electrolyte pairings by combining energetic affinity and electrolyte-surface morphology. The index includes two contributions: an energetic score, $S_\mathrm{energy}$, and an electrolyte morphology score, $S_{\mathrm{morph,ely}}$.

The energetic score is calculated from two interfacial descriptors: the thermodynamic work of adhesion, $W_{12}$, and the estimated interfacial energy, $\gamma_{12}$. High $W_{12}$ values are considered favourable because they indicate a stronger thermodynamic tendency toward adhesion, while low $\gamma_{12}$ values indicate a more energetically favourable interface. For each electrode/electrolyte pair, the two descriptors are normalized over the full set of investigated combinations as:
\begin{equation}
S_W=\frac{W_{12}-W_{12,\min}}{W_{12,\max}-W_{12,\min}},
\end{equation}
\begin{equation}
S_\gamma=\frac{\gamma_{12,\max}-\gamma_{12}}{\gamma_{12,\max}-\gamma_{12,\min}}.
\end{equation}

The energetic score is then defined as:
\begin{equation}
S_\mathrm{energy}=0.55S_W+0.45S_\gamma.
\end{equation}
A slightly higher weight is assigned to $W_{12}$ because it directly represents the thermodynamic tendency of the two surfaces to interact, while $\gamma_{12}$ is used as a complementary penalty for less favourable interfaces.

The electrolyte morphology score is calculated from a normalized topographic severity parameter, $M_\mathrm{ely}$, based on selected ISO 25178 descriptors. These descriptors are chosen because they are related to incomplete coating, valley formation, local non-contact and surface irregularity. For each electrolyte and each roughness parameter $x_j$, the normalized penalty is defined as:
\begin{equation}
P_j=\frac{x_j-x_{j,\min}}{x_{j,\max}-x_{j,\min}},
\end{equation}
where $P_j=0$ corresponds to the most favourable morphology and $P_j=1$ to the most critical one within the investigated electrolyte set.

The overall morphology severity is calculated as:
\begin{equation}
M_\mathrm{ely}=0.10P_{S_a}+0.20P_{S_z}+0.30P_{S_v}+0.38P_{|S_{ku}-3|}+0.02P_{S_{dq}}.
\end{equation}
The selected terms account for average roughness, maximum peak-to-valley height, valley depth, deviation from a near-Gaussian height distribution, extreme topographic events and mean surface slope. The morphology score is then obtained by inverting and rescaling the severity parameter:
\begin{equation}
S_{\mathrm{morph,ely}}=S_{\min}+(S_{\max}-S_{\min})\frac{M_{\max}-M_\mathrm{ely}}{M_{\max}-M_{\min}},
\end{equation}
with $S_{\min}=0.74$ and $S_{\max}=0.99$. These values correspond respectively to the most critical and most favourable electrolyte morphology within the investigated set.

The electrode free-surface morphology is not included in the index. Although it provides useful information on coating quality and homogeneity, it does not necessarily represent the buried electrode/electrolyte interface formed after deposition, drying and thermal treatment. The morphology term therefore refers only to the electrolyte surface, which is the actual substrate onto which the electrode layer is deposited.

The final compatibility index is defined as:
\begin{equation}
I=0.5S_\mathrm{energy}+0.5S_{\mathrm{morph,ely}}.
\end{equation}
In this report, energetic affinity and electrolyte morphology are assigned equal weight. This choice reflects the absence of a direct validation dataset at this stage. The coefficients should be calibrated in future work against experimental indicators such as slurry spreading, coating continuity, cross-sectional microscopy, adhesion or delamination tests, thermal cycling, EIS-derived polarization resistance, ASR and electrical contact resistance.

The index $I$ is therefore interpreted as a relative screening metric for the present material set. It is not an intrinsic material property, a universal physical law or a direct predictor of electrochemical performance. Rather, it is used to identify electrode/electrolyte pairings that combine favourable surface energetics with an electrolyte morphology suitable for continuous electrode deposition and active-area formation.

\section{Results and Discussion}

\subsection{Surface free energy and wetting behaviour}

\begin{figure}[H]
    \centering
    \includegraphics[width=0.85\textwidth]{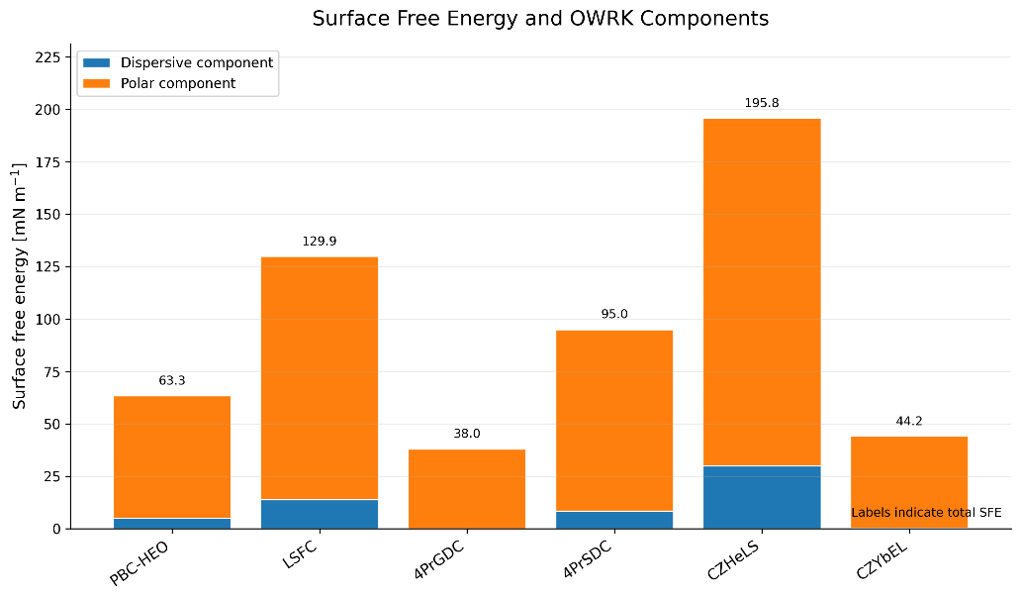}
    \caption{Total surface free energy and dispersive/polar OWRK contributions.}
    \label{fig:sfe}
\end{figure}

The surface free energy values obtained from contact-angle measurements and OWRK analysis are summarized in Table~\ref{tab:sfe}, while the relative dispersive and polar contributions are shown in Fig.~\ref{fig:sfe}.

\begin{table}[H]
\centering
\small
\caption{Contact angles and surface free energy values for the analysed materials.}
\label{tab:sfe}
\resizebox{\textwidth}{!}{%
\begin{tabular}{llrrrrr}
\toprule
Material & Role & Water CA [deg] & Glycerol CA [deg] & Total SFE [mN/m] & Dispersive SFE [mN/m] & Polar SFE [mN/m] \\
\midrule
PBC-HEO & Electrode & 38.19 & 48.05 & 63.34 & 5.04 & 58.30 \\
LSFC & Electrode & 50.78 & 88.44 & 129.88 & 13.99 & 115.89 \\
4PrGDC & Electrolyte & 77.69 & 89.75 & 38.03 & 0.00 & 38.03 \\
4PrSDC & Electrolyte & 63.08 & 92.84 & 95.01 & 8.44 & 86.57 \\
CZHeLS & Electrolyte & 34.56 & 87.93 & 195.75 & 29.99 & 165.77 \\
CZYbEL & Electrolyte & 76.73 & 91.79 & 44.19 & 0.30 & 43.89 \\
\bottomrule
\end{tabular}%
}
\end{table}

The first relevant result is that all investigated materials exhibit a dominant polar contribution to the total SFE. This behaviour is consistent with the oxide nature of the materials, where ionic, dipolar, acid-base and defect-related surface interactions are expected to play a major role.

Among the electrode materials, LSFC shows a significantly higher total SFE than PBC-HEO. This difference is mainly associated with the polar component, indicating a stronger tendency of LSFC toward specific surface interactions. Among the electrolytes, CZHeLS exhibits the highest total SFE, whereas 4PrGDC and CZYbEL are characterized by lower SFE values and almost negligible dispersive components.

A high total SFE does not automatically imply a better electrochemical interface. Charge transfer in SOFC/ReSOC electrodes depends on electrode kinetics, phase percolation, microstructure, porosity, temperature, atmosphere and the effective continuity of the electrode/electrolyte contact. Therefore, the SFE values in Table~\ref{tab:sfe} should be interpreted as descriptors of surface affinity and deposition tendency, not as direct predictors of ASR, ECR or charge-transfer resistance.

\begin{figure}[H]
    \centering
    \includegraphics[width=0.8\textwidth]{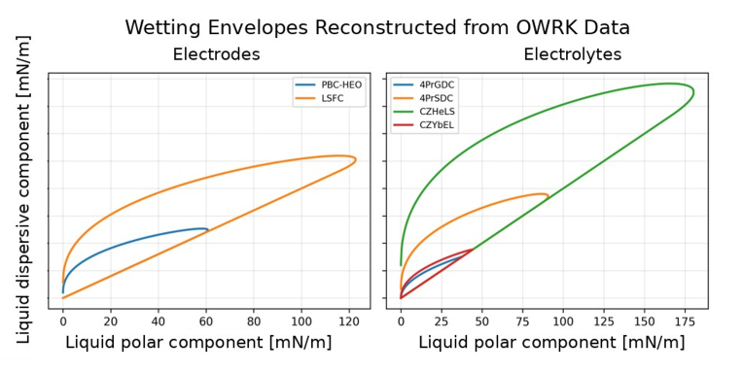}
    \caption{Wetting envelopes reconstructed from the OWRK method. The curves indicate the region of polar/dispersive liquid surface-tension composition compatible with surface wettability.}
    \label{fig:wetting}
\end{figure}

The wetting envelopes reported in Fig.~\ref{fig:wetting} provide a more process-oriented interpretation of the SFE data. The broader envelopes of LSFC and CZHeLS indicate a wider range of liquid polar/dispersive combinations that may wet these surfaces. This is relevant for electrode deposition, because the electrode layer is generally applied as a liquid suspension or slurry onto the electrolyte. In this sense, Fig.~\ref{fig:wetting} should be read not only as a wetting map, but also as a preliminary guide for selecting solvent/slurry formulations compatible with each electrolyte surface.

\subsection{Surface topography and role of electrolyte morphology}

\begin{figure}[H]
    \centering
    \includegraphics[width=0.78\textwidth]{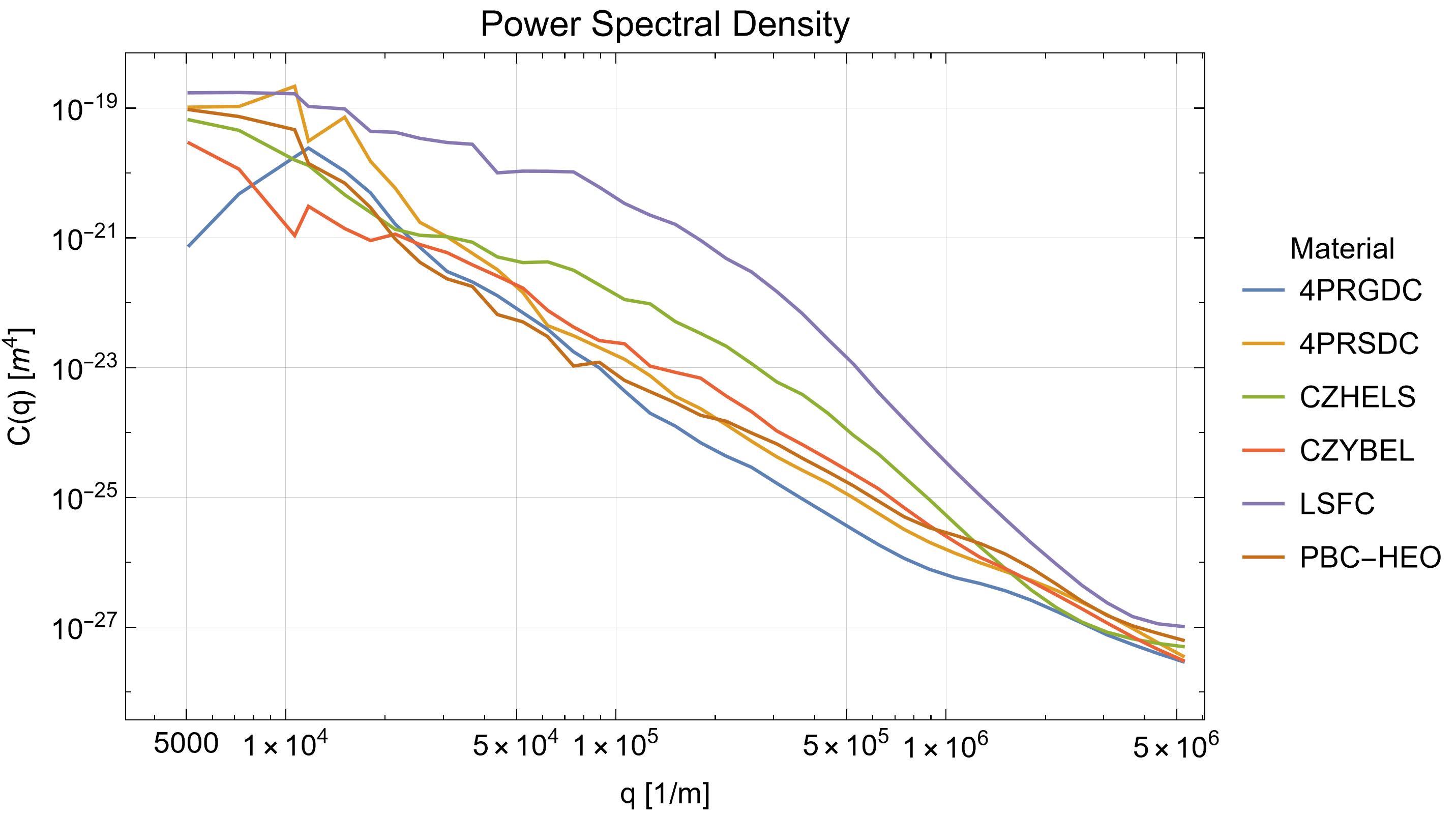}
    \caption{Power spectral density of the surface roughness.}
    \label{fig:psd}
\end{figure}

The surface topography results are summarized in Table~\ref{tab:roughness}, while the power spectral densities of the surface roughness are shown in Fig.~\ref{fig:psd}.

\begin{table}[H]
\centering
\small
\caption{ISO 25178-2 roughness parameters of the analysed electrode and electrolyte materials.}
\label{tab:roughness}
\resizebox{\textwidth}{!}{%
\begin{tabular}{llrrrrrrr}
\toprule
Material & Role & $S_a$ [$\mu$m] & $S_q$ [$\mu$m] & $S_z$ [$\mu$m] & $S_v$ [$\mu$m] & $S_{sk}$ & $S_{ku}$ & $S_{dq}$ \\
\midrule
PBC-HEO & Electrode & 8.59 & 11.00 & 189.00 & 122.00 & -0.111 & 6.33 & 1.38 \\
LSFC & Electrode & 23.70 & 30.10 & 227.00 & 149.00 & -1.24 & 4.39 & 4.04 \\
4PrGDC & Electrolyte & 4.03 & 5.23 & 37.50 & 19.20 & -0.252 & 3.54 & 0.956 \\
4PrSDC & Electrolyte & 6.97 & 8.92 & 52.60 & 27.90 & 0.0396 & 2.99 & 1.04 \\
CZHeLS & Electrolyte & 6.09 & 8.54 & 93.70 & 67.80 & -1.94 & 8.64 & 1.55 \\
CZYbEL & Electrolyte & 3.42 & 5.64 & 66.10 & 49.70 & -3.42 & 23.10 & 1.15 \\
\bottomrule
\end{tabular}%
}
\end{table}

The comparison confirms that the investigated materials differ not only in average roughness, but also in the severity of valleys, height-distribution asymmetry and extreme topographic events. The parameters $S_a$ and $S_q$ describe the average amplitude of roughness, but they are not sufficient to assess interface-forming ability. Parameters such as $S_z$, $S_v$, $S_{sk}$, $S_{ku}$ and $S_{dq}$ are more informative for interface design because they reveal the presence of deep valleys, isolated peaks, skewed height distributions and steep local slopes. These features may control whether the electrode precursor can conform to the electrolyte surface or whether micro-gaps and non-covered regions remain after deposition.

As shown in Table~\ref{tab:roughness}, the electrodes generally display higher roughness than the electrolytes, with LSFC being the most severe surface in terms of $S_a$, $S_q$ and $S_{dq}$. However, for the purpose of electrode/electrolyte interface screening, the morphology of the electrolyte is more relevant than the morphology of the electrode free surface. This is because, in the considered electrolyte-supported configuration, the electrode is deposited directly onto the electrolyte substrate. Therefore, the electrolyte surface is the topographical boundary condition that governs spreading, valley filling and buried-interface continuity.

The roughness measured on separately prepared electrode specimens does not necessarily represent the morphology of the buried electrode/electrolyte interface after deposition, drying and firing. For this reason, the electrode morphology is excluded from the final compatibility index, while the electrolyte morphology is retained as the relevant morphological descriptor.

\subsection{Energy descriptors and compatibility index}

\begin{figure}[H]
    \centering
    \includegraphics[width=\textwidth]{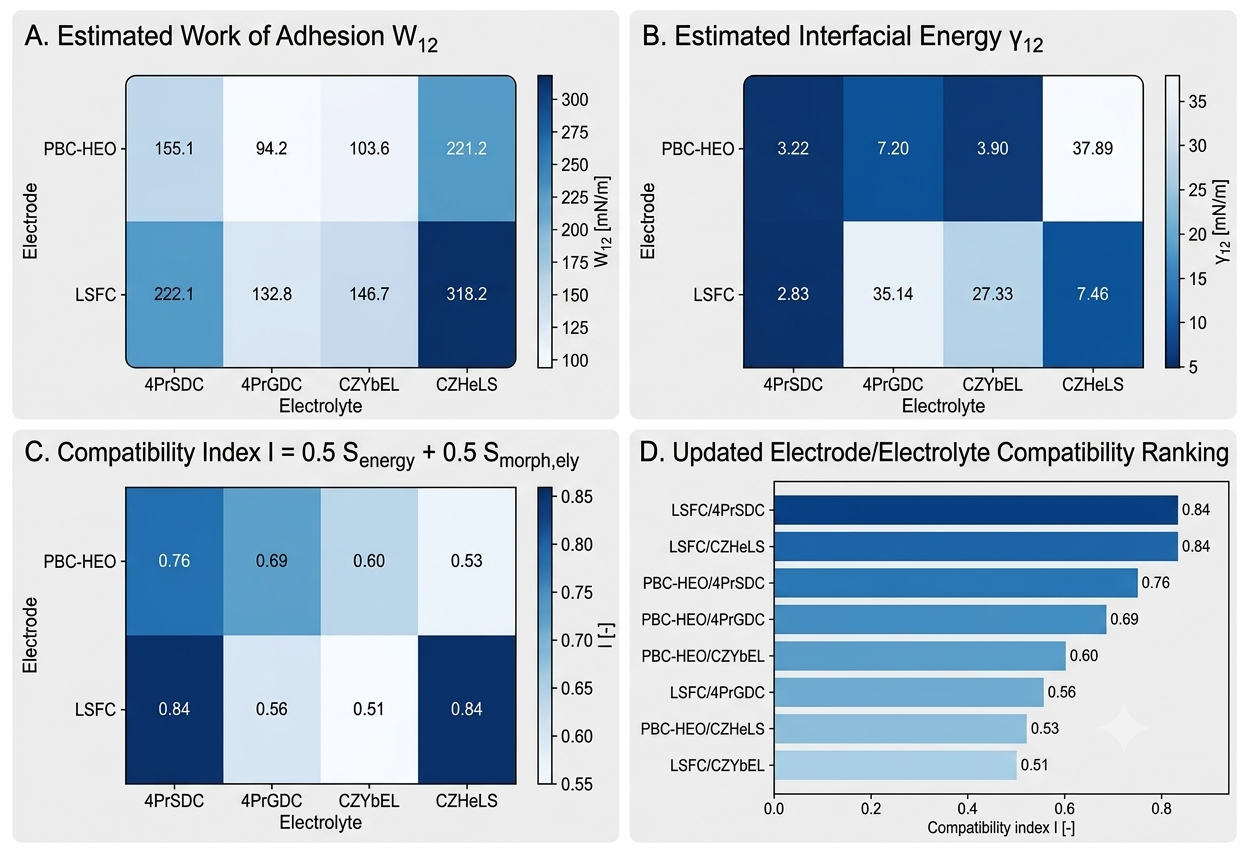}
    \caption{Compatibility maps for the eight electrode/electrolyte pairs: (A) estimated work of adhesion $W_{12}$; (B) estimated interfacial energy $\gamma_{12}$; (C) compatibility index $I=0.5S_\mathrm{energy}+0.5S_{\mathrm{morph,ely}}$; (D) resulting compatibility ranking.}
    \label{fig:heatmaps}
\end{figure}

The energetic and morphology descriptors are reported in Table~\ref{tab:compatibility}. The corresponding heatmaps of the estimated thermodynamic work of adhesion and interfacial energy are shown in Fig.~\ref{fig:heatmaps}.

\begin{table}[H]
\centering
\small
\caption{Compatibility matrix: energetic descriptors, morphological score and integrated index.}
\label{tab:compatibility}
\resizebox{\textwidth}{!}{%
\begin{tabular}{rllrrrrr}
\toprule
Rank & Electrode & Electrolyte & $W_{12}$ [mN/m] & $\gamma_{12}$ [mN/m] & $S_\mathrm{energy}$ & $S_{\mathrm{morph,ely}}$ & $I$ \\
\midrule
1 & LSFC & 4PrSDC & 222.06 & 2.83 & 0.76 & 0.92 & 0.84 \\
2 & LSFC & CZHeLS & 318.17 & 7.46 & 0.94 & 0.74 & 0.84 \\
3 & PBC-HEO & 4PrSDC & 155.13 & 3.22 & 0.59 & 0.92 & 0.755 \\
4 & PBC-HEO & 4PrGDC & 94.17 & 7.20 & 0.39 & 0.99 & 0.69 \\
5 & PBC-HEO & CZYbEL & 103.63 & 3.90 & 0.46 & 0.75 & 0.605 \\
6 & LSFC & 4PrGDC & 132.77 & 35.14 & 0.13 & 0.99 & 0.56 \\
7 & PBC-HEO & CZHeLS & 221.20 & 37.89 & 0.31 & 0.74 & 0.525 \\
8 & LSFC & CZYbEL & 146.74 & 27.33 & 0.26 & 0.75 & 0.505 \\
\bottomrule
\end{tabular}%
}
\end{table}

The heatmap in Fig.~\ref{fig:heatmaps}A shows that the highest work of adhesion is obtained for LSFC/CZHeLS. This very high value reflects the strong polar character of both surfaces and indicates a strong thermodynamic tendency toward interaction. LSFC/4PrSDC also shows a high work of adhesion, while the lowest values are found for combinations involving 4PrGDC with PBC-HEO and LSFC, reflecting weaker energetic affinity.

The interfacial energy heatmap in Fig.~\ref{fig:heatmaps}B provides a complementary view. Lower values indicate more energetically favourable interfaces. From this perspective, LSFC/4PrSDC and PBC-HEO/4PrSDC are particularly favourable. PBC-HEO/CZYbEL also shows a low estimated interfacial energy.

An important point emerges by comparing Fig.~\ref{fig:heatmaps}A and Fig.~\ref{fig:heatmaps}B. A high $W_{12}$ does not always correspond to a low $\gamma_{12}$. For example, PBC-HEO/CZHeLS has a relatively high adhesion work but also the highest estimated interfacial energy. This means that the individual surface energies alone are not sufficient: what matters is the balance between the polar and dispersive components of the two surfaces. This justifies the use of a normalized energetic score rather than a ranking based on a single descriptor.

The heatmap of the compatibility index is shown in Fig.~\ref{fig:heatmaps}C. The ranking derived from the same index is shown in Fig.~\ref{fig:heatmaps}D. The updated ranking identifies two leading combinations, LSFC/4PrSDC and LSFC/CZHeLS, both with $I=0.84$.

\begin{figure}[H]
    \centering
    \includegraphics[width=0.78\textwidth]{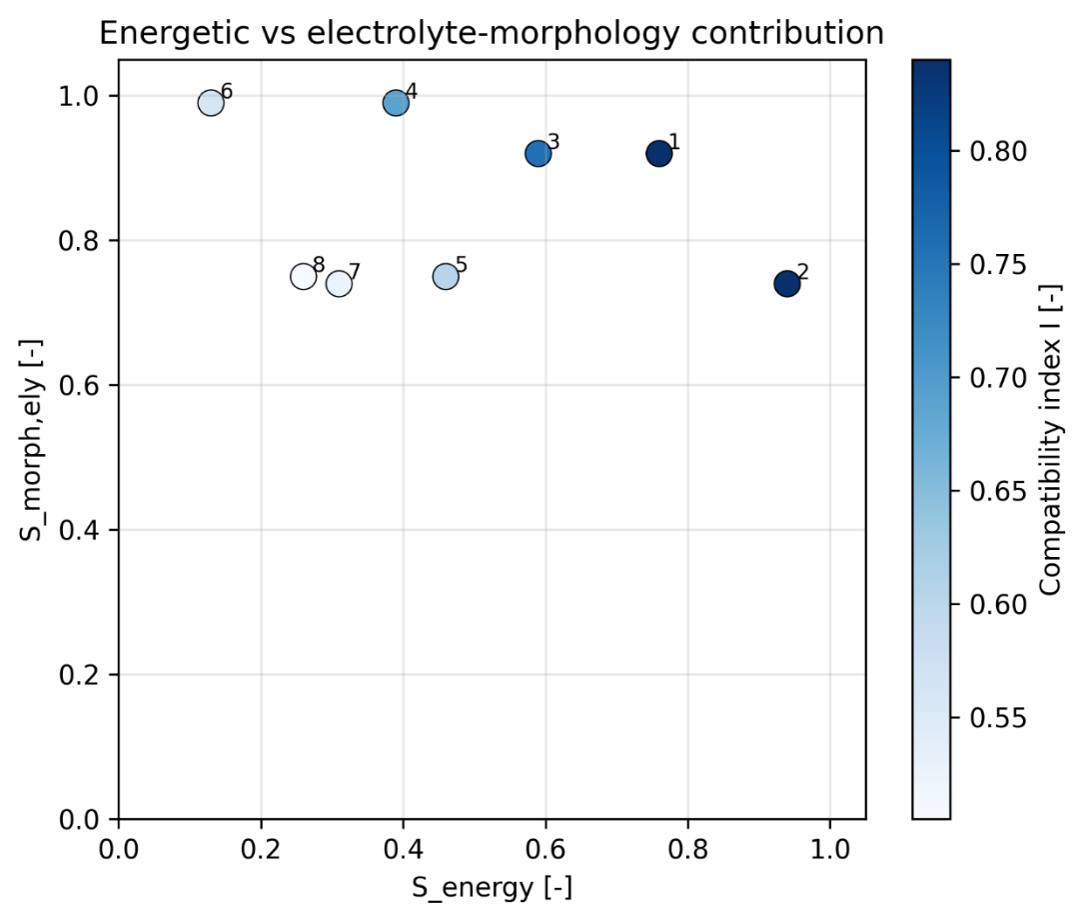}
    \caption{Energy-morphology representation of the compatibility score.}
    \label{fig:energy_morphology}
\end{figure}

Figure~\ref{fig:energy_morphology} provides a two-dimensional interpretation of the compatibility index by separating the energetic contribution, $S_\mathrm{energy}$, from the electrolyte morphology contribution, $S_{\mathrm{morph,ely}}$. This representation is useful because pairs with similar final compatibility index can originate from different balances between surface energetics and morphology. The most favourable region is the upper-right part of the plot, where high energetic compatibility is combined with a favourable electrolyte morphology. In this region, the interface is expected to be less penalized both by thermodynamic mismatch and by geometrical limitations during electrode deposition.

\section{Conclusions}

This work proposes a surface-based screening framework for electrode/electrolyte compatibility in SOFC/ReSOC materials, combining OWRK surface free energy, wetting behaviour and electrolyte morphology.

The results show that interface compatibility cannot be inferred from total surface free energy alone. A reliable assessment requires the combined evaluation of energetic affinity, through $W_{12}$ and $\gamma_{12}$, and electrolyte topography, which controls wetting, spreading and buried-interface continuity during electrode deposition.

For this reason, the final compatibility index $I$ is defined excluding the electrode free-surface morphology, which is not necessarily representative of the buried electrode/electrolyte interface.

The highest-ranking pairs are LSFC/CZHeLS and LSFC/4PrSDC, both with $I=0.84$, but with different meanings: LSFC/CZHeLS is mainly driven by high energetic affinity, whereas LSFC/4PrSDC provides the most balanced combination of surface energetics and electrolyte morphology. PBC-HEO/4PrSDC also emerges as a promising compromise.

The proposed matrix should be interpreted as a screening tool, not as a direct predictor of ASR, ECR or charge-transfer resistance. Its main value is to identify experimentally testable hypotheses on interface formation and active-area development. Future validation should include EIS, ASR/ECR measurements, cross-sectional microstructural analysis and adhesion or delamination tests.

\section*{Acknowledgements}

The authors gratefully acknowledge the financial support of the project BacForH2, Cod. RSH2A\_000038, funded under the Research and Development Programme for Hydrogen within the PNRR -- M2C2, Investment Line 3 (CUP: F57G25000330006).


\begin{thebibliography}{99}
\bibitem{minh1993} N. Q. Minh, ``Ceramic fuel cells,'' \textit{Journal of the American Ceramic Society}, 76, 563--588, 1993.
\bibitem{steele2001} B. C. H. Steele and A. Heinzel, ``Materials for fuel-cell technologies,'' \textit{Nature}, 414, 345--352, 2001.
\bibitem{singhal2000} S. C. Singhal, ``Advances in solid oxide fuel cell technology,'' \textit{Solid State Ionics}, 135, 305--313, 2000.
\bibitem{stambouli2002} A. B. Stambouli and E. Traversa, ``Solid oxide fuel cells (SOFCs): a review of an environmentally clean and efficient source of energy,'' \textit{Renewable and Sustainable Energy Reviews}, 6, 433--455, 2002.
\bibitem{kakac2007} S. Kaka\c{c}, A. Pramuanjaroenkij and X. Y. Zhou, ``A review of numerical modeling of solid oxide fuel cells,'' \textit{International Journal of Hydrogen Energy}, 32, 761--786, 2007.
\bibitem{hajimolana2011} S. A. Hajimolana, M. A. Hussain, W. M. A. W. Daud, M. Soroush and A. Shamiri, ``Mathematical modeling of solid oxide fuel cells: a review,'' \textit{Renewable and Sustainable Energy Reviews}, 15, 1893--1917, 2011.
\bibitem{golkhatmi2022} S. Zarabi Golkhatmi, M. I. Asghar and P. D. Lund, ``A review on solid oxide fuel cell durability: latest progress, mechanisms, and study tools,'' \textit{Renewable and Sustainable Energy Reviews}, 161, 112339, 2022.
\bibitem{chen2011} K. Chen and S. P. Jiang, ``Review: failure mechanisms of solid oxide fuel cells,'' \textit{International Journal of Hydrogen Energy}, 36, 10541--10549, 2011.
\bibitem{laguna2012} M. A. Laguna-Bercero, ``Recent advances in high temperature electrolysis using solid oxide fuel cells: a review,'' \textit{Journal of Power Sources}, 203, 4--16, 2012.
\bibitem{ebbesen2014} S. D. Ebbesen, S. H. Jensen, A. Hauch and M. B. Mogensen, ``High temperature electrolysis in alkaline cells, solid proton conducting cells, and solid oxide cells,'' \textit{Chemical Reviews}, 114, 10697--10734, 2014.
\bibitem{atkinson2004} A. Atkinson et al., ``Advanced anodes for high-temperature fuel cells,'' \textit{Nature Materials}, 3, 17--27, 2004.
\bibitem{simwonis2000} D. Simwonis, F. Tietz and D. St\"over, ``Nickel coarsening in annealed Ni/8YSZ anode substrates for solid oxide fuel cells,'' \textit{Solid State Ionics}, 132, 241--251, 2000.
\bibitem{hauch2008} A. Hauch, S. D. Ebbesen, S. H. Jensen and M. Mogensen, ``Solid oxide electrolysis cells: microstructure and degradation of the Ni/YSZ electrode,'' \textit{Journal of The Electrochemical Society}, 155, B1184--B1193, 2008.
\bibitem{jiang2008} S. P. Jiang, ``Development of lanthanum strontium manganite perovskite cathode materials of solid oxide fuel cells: a review,'' \textit{Journal of Materials Science}, 43, 6799--6833, 2008.
\bibitem{shao2004} Z. Shao and S. M. Haile, ``A high-performance cathode for the next generation of solid-oxide fuel cells,'' \textit{Nature}, 431, 170--173, 2004.
\bibitem{adler2004} S. B. Adler, ``Factors governing oxygen reduction in solid oxide fuel cell cathodes,'' \textit{Chemical Reviews}, 104, 4791--4843, 2004.
\bibitem{yokokawa2008} H. Yokokawa, H. Tu, B. Iwanschitz and A. Mai, ``Fundamental mechanisms limiting solid oxide fuel cell durability,'' \textit{Journal of Power Sources}, 182, 400--412, 2008.
\bibitem{costamagna1998} P. Costamagna, P. Costa and V. Antonucci, ``Micro-modelling of solid oxide fuel cell electrodes,'' \textit{Electrochimica Acta}, 43, 375--394, 1998.
\bibitem{virkar2000} A. V. Virkar, J. Chen, C. W. Tanner and J.-W. Kim, ``The role of electrode microstructure on activation and concentration polarizations in solid oxide fuel cells,'' \textit{Solid State Ionics}, 131, 189--198, 2000.
\bibitem{wilson2006} J. R. Wilson et al., ``Three-dimensional reconstruction of a solid-oxide fuel-cell anode,'' \textit{Nature Materials}, 5, 541--544, 2006.
\bibitem{jorgensen2010} P. S. J\o rgensen, K. V. Hansen, R. Larsen and J. R. Bowen, ``High accuracy interface characterization of three phase material systems in three dimensions,'' \textit{Journal of Power Sources}, 195, 8168--8176, 2010.
\bibitem{fu2015} Y. Fu et al., ``Heterogeneous electrocatalysis in porous cathodes of solid oxide fuel cells,'' \textit{Journal of The Electrochemical Society}, 162, F613--F621, 2015.
\bibitem{brandon2006} N. P. Brandon and D. J. L. Brett, ``Engineering porous materials for fuel cell applications,'' \textit{Philosophical Transactions of the Royal Society A}, 364, 147--159, 2006.
\bibitem{kishimoto2011} M. Kishimoto, H. Iwai, M. Saito and H. Yoshida, ``Quantitative evaluation of transport properties of solid oxide fuel cell porous anodes based on focused ion beam and scanning electron microscope reconstruction,'' \textit{Journal of Power Sources}, 196, 4555--4563, 2011.
\bibitem{chen2011pf} H.-Y. Chen et al., ``Ni coarsening in the three-phase solid oxide fuel cell anode: a phase-field simulation study,'' \textit{Journal of Power Sources}, 196, 1333--1337, 2011.
\bibitem{wang2015crack} X. Wang, Z. Chen and A. Atkinson, ``Crack formation in ceramic films used in solid oxide fuel cells,'' \textit{Journal of the European Ceramic Society}, 35, 391--397, 2015.
\bibitem{beckel2007} D. Beckel et al., ``Electrochemical performance of LSCF based thin film cathodes prepared by spray pyrolysis,'' \textit{Solid State Ionics}, 178, 407--415, 2007.
\bibitem{selcuk2001} A. Sel\c{c}uk, G. Merere and A. Atkinson, ``The influence of electrodes on the strength of planar zirconia solid oxide fuel cells,'' \textit{Journal of Materials Science}, 36, 1173--1182, 2001.
\bibitem{boukamp1995} B. A. Boukamp, ``A linear Kronig--Kramers transform test for immittance data validation,'' \textit{Journal of The Electrochemical Society}, 142, 1885--1894, 1995.
\bibitem{mogensen2019} M. Mogensen et al., ``Reversible solid oxide cells,'' \textit{Clean Energy}, 3, 175--201, 2019.
\bibitem{kim2020} Y.-D. Kim et al., ``Degradation studies of ceria-based solid oxide fuel cells at intermediate temperature under various load conditions,'' \textit{Journal of Power Sources}, 452, 227758, 2020.
\bibitem{faes2012} A. Faes, A. Hessler-Wyser, A. Zryd and J. Van Herle, ``A review of redox cycling of solid oxide fuel cells anode,'' \textit{Membranes}, 2, 585--664, 2012.
\bibitem{griffith1921} A. A. Griffith, ``The phenomena of rupture and flow in solids,'' \textit{Philosophical Transactions of the Royal Society A}, 221, 163--198, 1921.
\bibitem{irwin1957} G. R. Irwin, ``Analysis of stresses and strains near the end of a crack traversing a plate,'' \textit{Journal of Applied Mechanics}, 24, 361--364, 1957.
\bibitem{wang2015fracture} X. Wang, F. He, Z. Chen and A. Atkinson, ``Porous LSCF/dense 3YSZ interface fracture toughness measured by single cantilever beam wedge test,'' preprint/article version, 2015.
\bibitem{young1805} T. Young, ``An essay on the cohesion of fluids,'' \textit{Philosophical Transactions of the Royal Society of London}, 95, 65--87, 1805.
\bibitem{fowkes1964} F. M. Fowkes, ``Attractive forces at interfaces,'' \textit{Industrial \& Engineering Chemistry}, 56, 40--52, 1964.
\bibitem{owens1969} D. K. Owens and R. C. Wendt, ``Estimation of the surface free energy of polymers,'' \textit{Journal of Applied Polymer Science}, 13, 1741--1747, 1969.
\bibitem{kaelble1970} D. H. Kaelble, ``Dispersion-polar surface tension properties of organic solids,'' \textit{Journal of Adhesion}, 2, 66--81, 1970.
\bibitem{rabel1971} W. Rabel, ``Einige Aspekte der Benetzungstheorie und ihre Anwendung auf die Untersuchung und Ver\"anderung der Oberfl\"acheneigenschaften von Polymeren,'' \textit{Farbe und Lack}, 77, 997--1005, 1971.
\bibitem{degennes1985} P. G. de Gennes, ``Wetting: statics and dynamics,'' \textit{Reviews of Modern Physics}, 57, 827--863, 1985.
\bibitem{wenzel1936} R. N. Wenzel, ``Resistance of solid surfaces to wetting by water,'' \textit{Industrial \& Engineering Chemistry}, 28, 988--994, 1936.
\bibitem{cassie1944} A. B. D. Cassie and S. Baxter, ``Wettability of porous surfaces,'' \textit{Transactions of the Faraday Society}, 40, 546--551, 1944.
\bibitem{quere2008} D. Qu\'er\'e, ``Wetting and roughness,'' \textit{Annual Review of Materials Research}, 38, 71--99, 2008.
\bibitem{dubey2022} B. P. Dubey, A. Sahoo and Y. Sharma, ``Tailoring the surface energy and area surface resistance of solid-electrolyte polymer membrane for dendrite-free, high-performance, and safe solid-state Li-batteries,'' \textit{Journal of Power Sources}, 541, 231690, 2022.
\bibitem{banerjee2020} A. Banerjee, X. Wang, C. Fang, E. A. Wu and Y. S. Meng, ``Interfaces and interphases in all-solid-state batteries with inorganic solid electrolytes,'' \textit{Chemical Reviews}, 120, 6878--6933, 2020.
\bibitem{pervez2020} S. A. Pervez et al., ``Overcoming the interfacial limitations imposed by the solid--solid interface in solid-state batteries using ionic liquid-based interlayers,'' \textit{Small}, 16, 2000279, 2020.
\end{thebibliography}
\end{document}